\documentstyle[preprint,aps]{revtex}
\begin{document}
\draft
\preprint{hep-th/9509135, SNUTP/95-91}
\title{Classical and Quantum Mechanics of \\
 Non-Abelian Chern-Simons Particles }

\author{Phillial Oh\cite{poh}}
\address{Department of Physics,
Sung Kyun Kwan University,
Suwon 440-746,  KOREA}

\maketitle

\begin{abstract}
We investigate the classical and quantum properties
of a system of $SU(N)$ non-Abelian Chern-Simons (NACS)  particles.
After a brief introduction to the subject of NACS  particles, we
first  discuss about the symplectic structure of various $SU(N)$
coadjoint orbits which are the reduced phase space of
$SU(N)$ internal degrees of freedom or isospins.
A complete Dirac's constraint analysis is carried out
on each orbit and the Dirac bracket relations among
the isospin variables are calculated.
Then, the spatial degrees of freedom and
interaction with external gauge field are introduced by
considering the total reduced phase space which is
given by an associated bundle whose
fiber is one of the coadjoint orbits.
Finally, the theory is quantized by using the coherent state
method and various quantum mechanical properties are discussed
in this approach. In particular, a coherent state representation
of the   Knizhnik-Zamolodchikov equation is given and
possible solutions in this representation are  discussed.
\end{abstract}

\narrowtext

\newpage
\def\theequation{\arabic{section}.\arabic{equation}}
\section{Introduction}
\setcounter{equation}{0}
It is well known that  the angular
momentum is not quantized in two spatial dimensions
because the rotation group $SO(2)$ is Abelian and this leads to
one of the peculiar quantum mechanical properties of
physical systems in two spatial dimensions: the existence
of anyon and fractional spin and braid statistics \cite{wili,wilz}.
They found  many applications  to various areas of physics \cite{fort}
and in particular the anyon could be realized in the fractional
quantum Hall effect and perhaps in high temperature
superconductivity \cite{wili,pran}.

The notion of anyon can be generalized to non-Abelian
anyon, that is anyon with internal degrees of freedom
\cite{blok,oh1,oh2,bak1,kiml,oh3}. Especially in Ref. \cite{oh1},
the quantum mechanical model \cite{verl} of
non-relativistic $SU(2)$  non-Abelian
 Chern-Simons (NACS) particles which carry non-Abelian charges and
interact with each other through
the non-Abelian Chern-Simons terms \cite{dese,witt}  was derived from
a classical action principle and a detailed analysis of the model showed
that they lead to
the non-Abelian generalization \cite{froh} of fractional spin and
braid statistics. Later, a model of
 $SU(N)$ NACS particles  was constructed
by considering the internal degrees of freedom defined on
complex projective space $CP(N-1)$
 \cite{oh2}. This was generalized to an arbitrary group
 with invariant nonsingular metric\cite{bak1} and an equivalent
field-theoretic description  of NACS particles
was given \cite{bak1,kiml}.
Also, in Ref. \cite{oh3}, a Hamiltonian formalism of
$SU(N)$ NACS particles on complex projective space
was pursed by studying  the symplectic structure of the
reduced phase space of NACS particles which is given by an
associated bundle \cite{ster}.
The purpose of this paper is to extend the Hamiltonian analysis of
 the previous work \cite{oh3} to other possible symplectic
 manifolds with $SU(N)$ symmetry  and
perform a rigorous coherent state quantization \cite{klau}
of  resulting classical NACS particles.

We start by giving  a brief review of salient features of $SU(2)$
NACS particles theory to make this article
self-complete. Let us first consider the anyon case.
Anyons can be realized as  particles carrying both
 charge and magnetic flux and a possible quantum
mechanical model for them can be
 constructed \cite{fort} by considering a system of non-relativistic
charged point particles coupled with the Abelian
Chern-Simons gauge field \cite{hage}:
\begin{equation}
 L^\prime =
 \sum_\alpha {1 \over 2} m_\alpha \dot{\bf
q}_\alpha^2 +
\int d^2{\bf x} A_\mu({\bf x},t)j^\mu({\bf x},t)
+\frac{\theta}{2}\int
d^2 {\bf x} \epsilon^{\mu\nu\rho} A_\mu \partial_\nu
A_\rho
\label{lag}
\end{equation}
where
\begin{equation}
j^\mu({\bf x},t)\equiv (\rho({\bf x},t), {\bf j}({\bf x},t))
=\sum_\alpha e_\alpha\dot {\bf x}^\mu\delta^2({\bf x}-{\bf q}_\alpha).
\end{equation}
$e_\alpha$ ($\alpha=1,\cdots,N_p$) is the charge of each particle.
The Hamiltonian is given by
\begin{equation}
H^\prime = \sum_\alpha
{1\over 2 m_\alpha}\left( p^i_\alpha-e_\alpha A^{i}({\bf q}_\alpha)
\right)^2 -\int d^2{\bf x} A_0({\bf x},t)G({\bf x},t)
\end{equation}
where $G({\bf x},t)$ is the Gauss law constraints
\begin{equation}
-\theta B({\bf x},t)+\rho({\bf x},t)=0
\end{equation}
with the magnetic field  $B({\bf x},t)$.
The Gauss constraints can be solved in the Coulomb gauge
 $\nabla \cdot{\bf A}=0$ and we find
\begin{equation}
A_i({\bf x},t)=\frac{1}{2\pi\theta}\sum_\alpha
\frac{e_\alpha\epsilon_{ij}
({\bf x}-{\bf q_\alpha})_j}{\vert {\bf x}-{\bf q_\alpha}\vert^2}.
\end{equation}
This potential endows each particle with magnetic flux $\Phi_\alpha=
e_\alpha/\theta$ and provides
statistical interactions between anyons. Alternatively,
we can eliminate the interaction by a singular gauge transformation
\begin{equation}
\psi^\prime({\bf q}_1,\cdots,{\bf q}_{N_p})=
\prod_{\alpha<\beta}\exp \left[i\frac{ e_\alpha}{\pi\theta}
\Theta_{\alpha\beta}\right]\psi({\bf q}_1,\cdots,{\bf q}_{N_p})
\end{equation}
where $\Theta_{\alpha\beta}$ is the relative polar angle between particles
$\alpha$ and $\beta$.
In this case, the Hamiltonian becomes free but
the wave function $\psi^\prime$ is multi-valued
and this is the description of anyon
in the  so-called anyon gauge \cite{fort}.

In the non-Abelian case, we expect
\begin{equation}
H_0 = \sum_\alpha {1\over 2 m_\alpha}
\left(p^i_\alpha-A^{ai}({\bf q}_\alpha)
Q^a_\alpha\right)^2\label{mhami}
\end{equation}
where $A^a_\mu$ is the non-Abelian gauge field and $Q^a$ is the
generator of a non-Abelian gauge group $G$ with structure
constants $f_{abc}$'s:
$[Q^a_\alpha, Q^b_\beta] =
if^{ab}_{\ \ c}Q^c_\alpha\delta_{\alpha\beta}$.
The Gauss constraints would be
\begin{equation}
-\kappa B^a({\bf x},t)+\sum_\alpha Q^a_\alpha
\delta({\bf x}-{\bf q}_\alpha)=0
\end{equation}
for some constant $\kappa$.
It turns out the above Hamiltonian and Gauss constraints can be derived from
a classical action principle \cite{oh1,oh2}. It can be constructed in
terms of their spatial coordinates ${\bf q}_\alpha$'s
 and the isospin functions $Q^a_\alpha$'s which
transform under
the adjoint representation of the internal symmetry group.
Defining the isospin
functions directly on the reduced phase space which is $S^2$ for the
internal symmetry $SU(2)$,
\begin{equation}
Q^1_\alpha= J_\alpha \sin \theta_\alpha \cos\phi_\alpha,\quad Q^2_\alpha=
J_\alpha\sin \theta_\alpha \sin \phi_\alpha,\quad Q^3_\alpha =
J_\alpha\cos\theta_\alpha \label{iso}
\end{equation}
where $\theta_\alpha,\, \phi_\alpha$ are the coordinates
of the internal $S^2$ and $J_\alpha$ is a constant, one may write the
Lagrangian as \cite{oh1}
\[ L = \sum_\alpha\left({1 \over 2} m_\alpha \dot{\bf
q}_\alpha^2 +J_\alpha \cos \theta_\alpha \dot{\phi}_\alpha\right)
-\kappa\int
d^2 {\bf x} \,\epsilon^{\mu\nu\lambda} {\rm tr}\left(A_\mu \partial_\nu
A_\lambda +{2\over 3} A_\mu A_\nu A_\lambda\right) \]
\begin{equation}
+\int d^2{\bf x}\sum_\alpha \left(A^a_i(t,{\bf x}) \dot
q^i_\alpha -A^a_0(t, {\bf x})\right) Q^a_\alpha \delta ({\bf
x}-{\bf q}_\alpha).\label{lage}
\end{equation}
Here $\kappa=k/4\pi$, $k = {\rm integer}$, $A_\mu=A_\mu^a T^a$,
$[T^a, T^b] = -\epsilon^{abc} T^c$ and ${\rm tr} (T^a T^b) =
 -1/2 \delta_{ab}$.
The equations of motion from
the Lagrangian Eq. (\ref{lage}) contain Wong's equations\cite{wong}.

The above Gauss constraints can be solved  explicitly
in two gauge conditions.
The first one is the axial gauge\cite{kapu,bak1}
in which, for example, we set $A_1^a=0.$  The remaining $A_2^a$ field
becomes highly singular with strings attached to each source.
The less singular solutions can be obtained by performing
an analytic continuation of the gauge fields.
Introducing complex spatial coordinates, $z = x+ iy$, $\bar z
= x- iy$, $z_\alpha = q^1_\alpha + i q^2_\alpha$, $\bar z_\alpha =
q^1_\alpha
- i q^2_\alpha$, $A^a_z = {1\over 2} (A^a_1 - iA^a_2)$, $A^a_{\bar z} =
{1\over 2} (A^a_1 + iA^a_2)$, analytic continuation means that
$A^a_z$ and $A^a_{\bar z}$ are treated as independent variables
which is consistent with the coherent state quantization scheme\cite{fadd2}.
We  choose $A^a_{\bar z}=0$ as a gauge fixing condition
 which was called holomorphic gauge in Ref. \cite{oh1}.
The solution of the Gauss constraints
\begin{equation}
\Phi^a(z)=-\kappa \partial_{\bar z} A^a_z +\sum_\alpha
Q^a_\alpha\delta(z-z_\alpha)=0 \label{gauss}
\end{equation}
in holomorphic gauge turns out to be\cite{oh1}
\begin{equation}
A^a_z (z, \bar z) = {i\over 2\pi \kappa}\sum_\alpha  Q^a_\alpha
{1\over z -z_\alpha}+P(z)\label{sol}
\end{equation}
where $P(z)$ is an arbitrary holomorphic polynomial in $z$.
The further choice of $P(z)=0$, which is usually
known as  Knizhnik-Zamolodchikov (KZ) connection,
results in a quantum mechanical model \cite{verl,oh1} which  provides a
 unified framework for fractional spin, braid statistics and
KZ equation\cite{kniz}.

Substituting the above solution into the $N_p$ particle Hamiltonian
 Eq. (\ref{mhami}), we obtain
\begin{equation}
H=\sum_\alpha \frac{2}{m_\alpha}p_\alpha^z\left(p^{\bar z}_\alpha-
{i\over 2\pi \kappa}\sum_\beta \frac{ Q^a_\alpha  Q^a_\beta}
{ z_\alpha -z_\beta}\right).\label{prop}
\end{equation}
Quantum mechanically, the dynamics of the NACS particles are
 governed by the
operator version  ${\hat H}$ of the Hamiltonian Eq. (\ref{prop})
\cite{verl,oh1,oh2,bak1,kiml}
\begin{eqnarray}
 \hat {H}&=& -\sum_\alpha {1\over m_\alpha}\left(\nabla_{\bar
z_\alpha}\nabla_{z_\alpha}  +\nabla_{z_\alpha}\nabla_{\bar
z_\alpha}\right)\nonumber\\
\nabla_{z_\alpha}&=&{\partial\over \partial z_\alpha}  +{1\over 2\pi
\kappa}\left( \sum_{\beta\not=\alpha}
\hat Q^a_\alpha \hat Q^a_\beta {1\over
z_\alpha -z_\beta}+\hat Q^2_\alpha a_z (z_\alpha)\right)\label{ham}\\
\nabla_{\bar z_\alpha} &=&{\partial\over \partial \bar z_\alpha}
\nonumber
\end{eqnarray}
where $a_z (z_\alpha)=\lim_{z\rightarrow z_\alpha} 1/(z-z_\alpha)$
and the
isospin operators $\hat Q^a$'s satisfy the $SU(2)$
algebra, $[\hat Q^a_\alpha,\hat Q^b_\beta] =i\epsilon^{abc} \hat
Q^c_\alpha \delta_{\alpha\beta}$ upon quantizing the classical
Poisson bracket of isospin functions (\ref{iso}).
 The second term and the third term
in $\nabla_{z_\alpha}$ are
responsible for the non-Abelian statistics and the
exotic spins of the NACS particles respectively.
This can be seen  if
the wave function $\Psi_h$ for the NACS particles
in the holomorphic gauge is expressed as follows:
\begin{eqnarray}
\Psi_h(z_1,\dots,z_{N_p}) &=& U^{-1}(z_1,\dots,z_{N_p}) U^{-1}_{\rm s}
\Psi_a (z_1,\dots,z_{N_p}) \nonumber \\
U^{-1}_{\rm s} &=& \exp\left(-{1\over 2\pi \kappa}\sum_\alpha
\lim_{z\rightarrow z_\alpha}\int^z
{\hat Q^2_\alpha \over z-z_\alpha} dz\right)
\end{eqnarray}
where $U^{-1}(z_1,\dots,z_{N_p})$ satisfies
the KZ equation
\cite{kniz}
\begin{equation}
\left({\partial\over \partial z_\alpha}  + {1\over 2\pi
\kappa} \sum_{\beta\not=\alpha} \hat Q^a_\alpha \hat Q^a_\beta {1\over
z_\alpha -z_\beta}\right) U^{-1}(z_1,\dots,z_{N_p}) =0.\label{mkzeq}
\end{equation}

The KZ equation has a formal solution which is
expressed as a path ordered line integral in the
$N_p$-dimensional complex space
\begin{equation}
U^{-1}(z_1,\dots,z_{N_p}) = P \exp\left[-{1\over 2\pi\kappa} \int_\Gamma
\sum_\alpha dz^\prime_\alpha
\sum_{\beta\not=\alpha} \hat Q^a_\alpha \hat Q^a_\beta
\frac{1}{ z^\prime_\alpha -z^\prime_\beta}\right]
\end{equation}
where $\Gamma$ is a path in the
$N_p$-dimensional complex space with one end point fixed and the other being
$z_f = (z_1,\dots,z_{N_p})$.
Explicit evaluation \cite{oh1} of the above formal
expression gives the monodromy matrices or the braid matrices.
We see that $\Psi_a$ obeys the
the non-Abelian braid statistics due to
$U(z_1,\dots,z_{N_p})$ while the wave function
$\Psi_h$ obeys the ordinary statistics and the Hamiltonian for
NACS particles becomes free in terms of $\Psi_a (z_1,\dots,z_{N_p})$.
We also observe these particles carry fractional spin $2j_\alpha
(j_\alpha+1)/k$, because $\exp\left(-{1\over 2\pi \kappa}\lim_{z\rightarrow
z_\alpha}\int^z {\hat Q^2_\alpha \over z-z_\alpha} dz\right)$ acquires a
non-trivial phase $-{\hat Q^2_\alpha \over \kappa} i =
-2\pi i\left({2j_\alpha
(j_\alpha+1)\over k}\right)$ under $2\pi$ rotation.
In analogy with the Abelian Chern-Simons particle theory
we may call $\Psi_a$ the NACS particle wave function in the anyon gauge.
Therefore we have two equivalent
descriptions for the NACS particles as in the case of the
Abelian Chern-Simons particles: in the holomorphic gauge and in the
anyon gauge. $U(z_1,\dots,z_{N_p})$ is the singular
and non-unitary transformation
function between the two gauges. It also defines an inner product in the
holomorphic gauge
\begin{equation}
<\Psi_1 |\Psi_2> = \int \prod_\alpha dz_\alpha d\bar z_\alpha
 \bar\Psi_1(z) U^\dagger (z) U(z) \Psi_2 (z).\label{inner}
\end{equation}
 The Hamiltonian (\ref{ham})
in the holomorphic gauge is  hermitian with respect to this inner product.
The detailed analysis of  the above quantum mechanical model
was performed in  Ref. \cite{oh1}.

In this paper, we generalize the above model
to a system of $SU(N)$  NACS particles and
quantize it using the coherent state quantization method \cite{klau}.
To describe the $SU(N)$ internal degrees of freedom, we first
have to identify the phase space for them.
The natural candidates are the coadjoint orbits of $SU(N)$ group
because they are symplectic manifolds \cite{kril} with
$SU(N)$ symmetry and thus can be considered as the reduced phase
space of generalized Hamiltonian dynamics \cite{arno}.
In fact, it can be shown that they are the reduced phase spaces of
the cotangent bundle
$T^*SU(N)$ by the method of symplectic reduction \cite{arno}.
Having identified the coadjoint orbits as the
reduced phase spaces by  symplectic reductions,
one could proceed to formulate the whole theory by coupling
with the spatial and external gauge degrees of
freedom in the Lagrangian method with the first order formalism.
In this approach, one writes down $SU(N)$ generalizations of the
$SU(2)$ isospin functions (\ref{iso}) and the Lagrangian  (\ref{lage})
on the $SU(N)$ coadjoint orbits. However, since not much
is known about the coordinatization of them, we will not
have explicit expressions of symplectic structure and
isospin functions with a few
exception, for example, like the case of $CP(N-1)$ \cite{oh2}.
This could pose a practical difficulty for this approach.
One could still pursue the Lagrangian formulation by
using the unreduced $SU(N)$ coordinates but with
many constraints corresponding to each orbit put
in  the Lagrangian by Lagrange multipliers.
This would make the theory look rather cumbersome.
Instead we perform the Hamiltonian analysis in this paper.

The outline of the paper is as follows.
We first consider various coadjoint orbits
of $SU(N)$ group and study the symplectic
structure of them. Total phase space
will be obtained by coupling the internal degrees of freedom with the spatial
degrees of freedom in the external gauge field.
We will not include the phase space of the gauge field itself
to make the presentation simple.
Then, we quantize the system by
the coherent state quantization method \cite{klau} and discuss about
the various quantum mechanical properties in this approach.
In Section 2, we start from symplectic reduction of the cotangent bundle
of $SU(N)$ group to identify the coadjoint orbits as the reduced
phase spaces of internal degrees of freedom and investigate
the symplectic structure of the most general coadjoint orbit of
$SU(N)$ group. The Dirac's constraint analysis is carried out
on each orbit. In Section 3, the symplectic structure on
the total phase space which is an associated  bundle is
explicitly given. This section is drawn mostly from Ref. \cite{oh3} and
included in this paper for completeness.
In Section 4, coherent state quantization method
is applied to the NACS particles system defined on
the reduced phase space of the associated bundle
of Section 3 and quantum mechanical properties are discussed. In particular,
we represent the KZ equation as a coherent state
differential equation and
 discuss about the possible solutions of KZ equation in
this method. Section 5 contains conclusion and discussion.

\def\theequation{\arabic{section}.\arabic{equation}}
\section{ Symplectic Reduction and $SU(N)$ Isospin}
\setcounter{equation}{0}

We start from the configuration space for
the  internal degrees of freedom which can be taken as
the group manifold $G$. To do analysis in a canonical approach, consider
$T^*G\cong G\times {\cal G}^*$, where ${\cal G}^*$
is the dual of the Lie algebra ${\cal G}$ of the
group G \cite{arno}.
A natural symplectic left group action on $T^*G$ can be defined as
\begin{eqnarray}
G\times (G\times {\cal G}^*)&\longrightarrow&
G\times {\cal G}^*\nonumber\\
(g, (h,a))&\mapsto& (gh,a).
\end{eqnarray}
Let us define the moment map
 $\rho: T^*G\rightarrow {\cal G}^*$ via
\begin{equation}
<X, \rho(m)>=m\left(\frac{d}{dt}\Big\vert_{t=0}
\exp tX\circ g\right)
\end{equation}
where $X\in {\cal G}$ and
$m\in T^*_gG$ is the linear map of ${\cal G}\rightarrow {\bf R}$.
The reduced phase space for isospin degrees of freedom can be obtained as the
quotient space $\rho^{-1}(x)/G_x$ which is well
defined for regular value of $x$.
Here  $G_x$ is the stabilizer group of the point  $ x\in {\cal G}^*.$
The above procedure is called a symplectic reduction.
It can be shown that the reduced
phase space is  naturally identifiable with
  the coadjoint orbit
${\cal O}_x\equiv G\cdot x\subset {\cal G}^*$
 \cite{arno}:
\begin{equation}
\rho^{-1}(x)/G_x\cong G/G_x\cong  G\cdot x.
\end{equation}

The same reduction can be achieved using the Dirac's constraint analysis.
According to Dirac \cite{dira}, in general,
there arise first class and second class
constraints in the reduction of the phase space. In our case,
 momentum maps associated with  $G_x$
are the first class constraints and the rest are second class.
To see this, let us separate the momentum map
$\rho_a(m)\equiv <T^a, \rho(m)>$ with $T^a$
being the generator of the Lie
algebra ${\cal G}$, $[T^a, T^b]=-f^{ab}_{\ \ c}
T^c$ with  $\mbox{Tr}(T^aT^b)=-1/2\delta_{ab}$
 into two groups:
$T^\alpha$'s  and $T^i$'s where $T^\alpha$'s belong to
 the Lie algebra of stabilizer subgroup $G_x$ and $T^i$'s are the rest.
The indices $\alpha, \beta, \cdots, i
,j, \cdots$ will be used repeatedly unless confusion arises.
Then we have the following:
\begin{equation}
[T^\alpha, T^\beta]=-f^{\alpha\beta}_{\ \ \gamma}T^\gamma,\ \
[T^\alpha, T^i]=-f^{\alpha i}_{\ \ j}T^j,\ \
[T^i, T^j]=-f^{ij}_{\ \ k}T^k-f^{ij}_{\ \ \alpha}T^\alpha.
\end{equation}
  We also have  the Lie algebra homomorphism on $T^*G$ \cite{arno}:
\begin{equation}
\{\rho_a, \rho_b\}=-f^{\ \ c}_{ab}\rho_c.\label{poisson}
\end{equation}
Let us define $x_a=(x_\alpha, x_i)$.
Then the constrained space $\rho^{-1}(x)$ is
a subspace of $T^*G$ given by the level set
$\rho_\alpha=x_\alpha, \ \rho_i=x_i$. We can rewrite
these  equations  in terms of the constraints
$\Gamma_a\equiv \rho_a-x_a\approx 0$.
Since the group $G_x$ is the stabilizer
group of the point $x\in {\cal G}^*$,
we have $\mbox{Ad}^*(T^\alpha)(x_a)
=0$ and it gives $f_{\alpha\beta}^{\ \ \gamma}x_\gamma=
f_{\alpha i}^{\ \ j}x_j=0$.
So  we  have the following constraints algebra:
\begin{eqnarray}
&\{&\Gamma_\alpha, \Gamma_\beta\}\approx -
f_{\alpha\beta}^{\ \ \gamma}\Gamma_\gamma,\
\{\Gamma_\alpha, \Gamma_i\}\approx-
f_{\alpha i}^{\ \ j}\Gamma_j,\nonumber\\
&\{&\Gamma_i, \Gamma_j\}\approx-f_{ij}^{\ \ k}\Gamma_k-
f_{ij}^{\ \ \alpha}\Gamma_\alpha-c_{ij}
\end{eqnarray}
where we have $c_{ij}=f_{ij}^{\ \ \alpha}x_\alpha
+f_{ij}^{\ \ k}x_k\neq 0$.
We see that $\Gamma_\alpha\approx 0$ are the first class constraints while
 $\Gamma_i\approx0$ are second class constraints.
Reduction to $\rho^{-1}(x)/G_x\cong G/G_x$ is
achieved with a suitable gauge choice corresponding
to the first class constraints $\Gamma_\alpha$'s.

Let us consider possible $SU(N)$ coadjoint orbits of type
${\cal O}_{\{n_1,n_2,\cdots,n_l\}}
\equiv SU(N)/SU(n_1)\times\cdots\times SU(n_l)\times U(1)^{l-1}$.
Here we have $\sum_{i=1}^ln_i=N$ and the rank of the subgroup
$H\equiv SU(n_1)\times \cdots\times SU(n_l)\times U(1)^{l-1}$
is equal to $N-1$.
It is well known that there is a natural symplectic structure
on the coadjoint orbits of a Lie group \cite{kril}.
They also have the complex structure inherited from
the complex representation
 of ${\cal O}_{\{n_1,n_2,\cdots,n_l\}}
 =SL(N, {\bf C})/P_{\{n_1,n_2,\cdots, n_l\}}$,
where  $SL(N, {\bf C})$ is the complexification
of $SU(N)$ and $P_{\{n_1,n_2,\cdots, n_l\}}$ is a parabolic subgroup
of  $SL(N, {\bf C})$ which is the subgroup
of block upper triangular matrices in the
$(n_1+n_2+\cdots +n_l)\times (n_1+n_2+\cdots +n_l)$ block decomposition.
 Borel subgroup $B$ corresponds $P_{\{1,1,\cdots, 1\}}$.
 Together with the symplectic structure, they
become K\"ahler manifolds.
Let us assume that the symplectic two form is given
in the local complex coordinate $(\bar \xi, \xi)$
by the K\"ahler form
\begin{equation}
\Omega=\sum_{i,j}\Omega_{ij}d \xi^i\wedge d\bar \xi^j\label{sympe}
\end{equation}
where $\Omega_{ij}$ can be expressed in terms
of K\"ahler potential $W$ by
\begin{equation}
\Omega_{ij}=i\partial_i\bar\partial_j W.
\end{equation}
Then the Poisson bracket can be defined via
\begin{equation}
\{f,g\}=\sum_{i,k}\Omega^{ki}\left(\frac{\partial
f}{\partial  \xi^k}\frac{\partial g}{\partial\bar \xi^i}-
\frac{\partial g}{\partial \xi^k}
\frac{\partial f}{\partial\bar \xi^i}\right)
\label{pbracket}
\end{equation}
where the inverse  $\Omega^{ki}$ satisfies
$\Omega_{ik}\Omega^{kj}=\delta_i^j$.

Isospin degrees of freedom on the coadjoint orbit
${\cal O}_{\{n_1,n_2,\cdots,n_l\}}$ can be defined as \cite{duval}
\begin{equation}
Q=\mbox{Ad}^*(x)g=gxg^{-1} \quad g\in SU(N)
\end{equation}
where $x=i\mbox{diag}(x_1,x_2,\cdots,x_N)$.
Here $\sum_{i=1}^Nx_i=0$ and we choose without loss of generality
 $x_1=x_2=\cdots x_{n_1}>
x_{n_1+1}=x_{n_1+2}=\cdots x_{n_2}>
\cdots =x_{n_l-1}=x_{n_l}$.
The restriction is $n_i\leq N-1$.
When $n_1=N-1, n_2=1$ or $n_1=1, n_2=N-1$, the orbit
corresponds to the minimal orbit which is a complex projective space $CP(N)$.
When $n_1=n_2=\cdots n_l=1$, it corresponds to the maximal orbit
which is a flag manifold.
It is convenient to express the element $g$ of $SU(N)$
by $N$ column vectors $(Z_1, Z_2,\cdots, Z_N),
\quad Z_p\in {\bf C}^N$ $(p,q=1,\cdots,N)$
 such that
 \begin{equation}
\bar Z_pZ_q=\delta_{pq}, \quad \mbox{det}
(Z_1, Z_2,\cdots, Z_N) =1.\label{cond}
\end{equation}

Let us consider the canonical one form $\theta$
\begin{equation}
\theta=\mbox{Tr}(xg^{-1}dg)=i\sum_{p=1}^{N}x_p \bar Z_p dZ_p.
\end{equation}
Using the second equation of (\ref{cond}), we find
\begin{equation}
\theta=i\sum_{p=1}^{N-1}J_p\bar Z_p dZ_p, \quad
J_p=x_1+\cdots +2x_p+\cdots +x_{N-1}\geq 0
\end{equation}
and the symplectic two-form $\Omega^\prime$:
\begin{equation}
\Omega^\prime=d\theta=-i\sum_{p=1}^{N-1}J_pd Z_p\wedge d\bar Z_p.
\end{equation}
Note the inequality $J_{n_1}>J_{n_2}>\cdots >J_{n_l}$ and  there
still exist constraints $\bar Z_pZ_q-\delta_{pq}\approx 0\
(p,q=1,\cdots,N-1)$.
Also $Q$ can be expressed as
\begin{equation}
Q=i\sum_{p=1}^{N-1}(J_pZ_p\bar Z_p-(J_p/N)I).
\label{isosp}
\end{equation}
We define the isospin functions $Q^a$'s by
\begin{equation}
Q^a= \mbox{Tr}(QT^a).\label{isof}
\end{equation}
In the case of $SU(2)$ example, we have
\begin{equation}
Q^1=-\frac{J}{2}(\xi^0\xi^{1*}+\xi^1\xi^{0*}),
Q^2=-i\frac{J}{2}(\xi^0\xi^{1*}-\xi^1\xi^{0*}),
Q^3=-\frac{J}{2}(\xi^0\xi^{0*}-\xi^1\xi^{1*})
\end{equation}
with the constraint $\vert \xi^0\vert^2+\vert \xi^1\vert^2=1$.
This is the Hopf map of $S^3\rightarrow S^2$.

To find the Poisson bracket relations among the isospin functions $Q^a$'s,
we define the canonical Poisson bracket relations by \cite{faj,alek2}
\begin{equation}
\{\bar Z^i_p, Z^j_q\}=(i/J_p)\delta_{pq}\delta^{ij}
\quad (p,q=1,\cdots N-1;i,j=1,\cdots,N)\label{canoni}
\end{equation}
with the constraints $\bar Z_pZ_q-\delta_{pq}\approx 0$.
We suppose that none of the $J_p$'s are equal to zero for the
time being. The case in which some of the
$J_p$'s are equal to zero will be considered shortly after.
 Let us divide
the constraints into
\begin{equation}
\Psi_p=\bar Z_pZ_p-1\approx 0,\quad
\Phi_{pq}=\bar Z_pZ_q\approx 0 \ (p\neq q).\label{conddd}
\end{equation}
Using Eq. (\ref{canoni}), we can check that the following
constraint algebra holds:
\begin{eqnarray}
\{\Psi_p,\Psi_q\}&\approx& 0\nonumber\\
\{\Psi_p,\Phi_{qr}\}&=&(i/J_p)(\delta_{pr}\Phi_{qp}-
\delta_{pq}\Phi_{pr})\nonumber\\
\{\Phi_{pq},\Phi_{rs}\}&=&(i/J_p)\delta_{ps}\bar Z_rZ_q
-(i/J_q)\delta_{qr}\bar Z_pZ_s.\label{constr}
\end{eqnarray}
We see that each of $\Psi_p\approx 0$ is a first class constraint.
Also from the third equation of (\ref{constr}), we deduce that
each of $\Phi_{n_i}\equiv\Phi_{pq}(\sum_{j=1}^{i-1}n_j
\leq p,q\leq \sum_{j=1}^{i}n_j; i=1,2,\cdots,l; n_0=0)$
is a first class constraint and the rest $\Phi_{pq}$'s are
second class ones. So there are $(N-1)+\sum_{i=1}^ln_i(n_i-1)$
first class constraints and $(N-1)(N-2)-\sum_{i=1}^ln_i(n_i-1)$
second class constraints. Note that the dimension of the reduced
phase space is $2N(N-1)-2[(N-1)+\sum_{i=1}^ln_i(n_i-1)]-
[(N-1)(N-2)-\sum_{i=1}^ln_i(n_i-1)]=N^2-\sum_{i=1}^ln_i^2$
which coincides with
the dimension of the coadjoint orbit ${\cal O}_{\{n_1,n_2,\cdots,n_l\}}
=SU(N)/SU(n_1)\times\cdots \times SU(n_l)\times U(1)^{l-1}$.
The first class constraints generate the subgroup
$SU(n_1)\times\cdots \times SU(n_l)\times U(1)^{l-1}$.
A Dirac's constraint analysis on
the maximal coadjoint orbit of dimensions $N^2-N$
with only first class constraints was
performed before \cite{alek2}.
We can eliminate the second class constraints all together by
using the Dirac bracket \cite{dira}:
\begin{equation}
\{f,g\}^*=\{f,g\}-\sum_{pqrs}^\prime\{f,\Phi_{pq}\}D^{-1}_{pq,rs}
\{\Phi_{rs},g\}\label{diracbra}
\end{equation}
with
\begin{equation}
D^{-1}_{pq,rs}\equiv i\frac{J_pJ_q}{J_q-J_p}\delta_{pq,rs}
\quad (p\neq q).
\end{equation}
Here  $\prime$ denotes that the first class constraints $\Phi_{n_i}$'s
do not appear in the sum and
 $\delta_{pq,rs}=\delta_{ps}\delta_{qr}.$
Using the expression for isospin functions (\ref{isof}), we find
\begin{equation}
\{Q^a,Q^b\}^*=-f^{ab}_{\ \ c}Q^c.
\end{equation}
We note that the above equation implies that the relation
$\{Q^a, Q^b\}=-f^{ab}_{\ \ c}Q^c$  results if one calculates directly
on the reduced phase space using the Eq. (\ref{pbracket}).

When some of the $J_p$'s are zero,
we assume that all the $Z_p$'s for which
$J_p=0$ can be eliminated by the Eq. (\ref{conddd}) in terms of
$Z_q$'s for which $J_q\neq 0$ and these variables do not appear in the
consequent analysis. This assumption is safe in view of the fact that
the internal degrees of freedom defined in the Eq. (\ref{isosp})
does not contain the variables $Z_p$'s for which $J_p=0$ and one does not
have to consider the Dirac brackets which contain these variables.
Note that the constraint algebra (\ref{constr}) is also restricted to
the constraints which do not contain these variables. A nice example
of the above procedure is the case of $CP(N-1)$. Consider
$x=i\mbox{diag}(N-1, -1,\cdots,-1)$ so that $J_1=N-1$ and $J_2=
\cdots=J_{N-1}=0$. This orbit is the $CP(N-1)=SU(N)/SU(N-1)\times U(1).$
According to our prescription, the only remaining variables are
$Z_1\in {\bf C}^N$ with the constraints $\Psi_1=\bar Z_1Z_1-1\approx 0$
which is obviously first class and generates a circle action.
 So the reduced phase space becomes
$S^{2N-1}/S^1$ which is another representation of $CP(N-1)$.
The constraint equation  $\Psi_1=\bar Z_1Z_1-1\approx 0$
with $Z^T=(\Xi_0, \Xi_1,\cdots, \Xi_{N-1})$ can be
solved explicitly in the gauge choice $\bar \Xi_0=\Xi_0 (\neq 0)$
with the results
\begin{equation}
 \Xi_0=\frac{1}{\sqrt{1+\vert\xi\vert^2}},\ \ \
\Xi_i=\frac{\xi^i}{\sqrt{1+\vert\xi\vert^2}}.
\label{slp}
\end{equation}
Here $\xi^i=\Xi_i/\Xi_0$ is the inhomogeneous coordinate for $CP(N-1)$.
The isospin function (\ref{isof}) is expressed as
\begin{equation}
Q^a(\xi,\bar\xi)=Ji\sum_{I,K=0}^{N-1}
{\bar \Xi}_IT^a_{IK}\Xi_K \quad (I=0,i)
\label{deff2}
\end{equation}
where Eq. (\ref{slp}) is substituted into the final expression.

\def\theequation{\arabic{section}.\arabic{equation}}
\section{Coupling with the spatial degree of freedom}
\setcounter{equation}{0}
Consider two dimensional configuration space $M$ which
we assume to be a plane.
(We present one particle case and later extend to $N_p$
particles in a straightforward manner.)
To account for the spatial degree of freedom and
external gauge field, let us consider
the principal $G=SU(N)$ bundle $P$ over $M$: $P\rightarrow M$.
The universal phase space for isospin particles in external
gauge field can be defined as  the direct product
$T^*P\times{\cal O}_{\{n_1,n_2,\cdots,n_l\}}$ \cite{wein}.
The left action of
$G$ on $P$ defined by $(g\cdot p)=p\cdot g^{-1}$ can be lifted to $T^*P$
and let us denote the momentum map for this action by $-\nu$.
Also let us denote the momentum map for the $G$ action on
${\cal O}_{\{n_1,n_2,\cdots,n_l\}}$ by $\mu$. Applying the
symplectic reduction procedure \cite{arno}, we consider the
constrained manifold $(-\nu+\mu)^{-1}(0)$ and dividing by
$G$, we get the reduced phase space $(-\nu+\mu)^{-1}(0)/G$ \cite{wein}.
When a connection on $P$ is chosen, the reduced phase space becomes
diffeomorphic to the associated bundle  ${\cal P}\equiv \bar P\times_G
{\cal O}_{\{n_1,n_2,\cdots,n_l\}}$
where  ${\bar P}\rightarrow M$ is the pull-back bundle
of the bundle  $P$
by the projection $\pi^\prime:T^*M\rightarrow M$.
${\cal P}$ is  Sternberg's reduced phase space \cite{ster}
and it can be shown that a given connection $\Theta$ on $P$
determines a unique symplectic structure on ${\cal P}$.

The essence of  Sternberg's reduced phase space is that there exist a
unique form $\Omega_{\Theta}$ on ${\cal P}$ such that
\begin{equation}
d\langle \Theta, Q \rangle+\pi^* \Omega=\bar\pi^*\Omega_{\Theta}
\end{equation}
where $\pi$ is the projection
map: $\bar P\times{\cal O}_{\{n_1,n_2,\cdots,n_l\}}
\rightarrow {\cal O}_{\{n_1,n_2,\cdots,n_l\}} $
 and $\bar\pi$ is the projection map
$\bar\pi:\bar P\times{\cal O}_{\{n_1,n_2,\cdots,n_l\}}\rightarrow {\cal P}$.
 $\Omega$ is the symplectic two form on the coadjoint orbit (\ref{sympe}).
It can be shown that the two form
 $\Omega_{\Theta}$ is closed and nondegenerate in the case
when $\bar P$ is the pull-back of the bundle  $ P$ as above
and the connection is the pull-back of a connection defined on $P$
\cite{ster}.
Denoting  $\tilde\omega$ for the pull-back of $\omega$ which is the
canonical symplectic structure defined on $T^*M$  to ${\cal P}$
via the projection onto $T^*M$,
we have a symplectic structure on ${\cal P}$ as
\begin{equation}
\Omega_T=\tilde\omega+\Omega_{\Theta}.
\end{equation}
When $M$ is a plane, $T^*M$ is contractible and every associated bundle
is trivial. So we have
\begin{equation}
{\cal P}= P\times_G{\cal O}_{\{n_1,n_2,\cdots,n_l\}}=
T^*M\times {\cal O}_{\{n_1,n_2,\cdots,n_l\}}.
\end{equation}
In fact, the above holds for arbitrary Riemann surfaces $M$ \cite{oh3}.
Hence we have $\tilde\omega=\omega$ and
\begin{eqnarray}
\Omega_T & = & \omega+
\sigma^*(d\langle \Theta, Q \rangle+\pi^* \Omega) \nonumber  \\
      & = & \omega +d(A^aQ^a)+\Omega
\end{eqnarray}
where $\sigma$ is the cross section
:$ P\times_G{\cal O}_{\{n_1,n_2,\cdots,n_l\}}
\rightarrow  P\times{\cal O}_{\{n_1,n_2,\cdots,n_l\}}$ and
we used $\sigma^*\Theta=A$, the gauge field one form on $M$.
 Notice that $\omega+\Omega$ is not
gauge invariant. We must have Sternberg's two form
 $d\langle \Theta, Q \rangle$
to achieve the gauge invariance.
 Physically, this term describes the interaction
between isospin charge and the external gauge field.

Now, we explicitly calculate the
symplectic structure on ${\cal P}$
and prove the minimal substitutions for the non-Abelian case.
We start from the two form on ${\cal P}$ given by
\begin{equation}
\Omega_T=dp_i\wedge dq^i+ d(A^a_i Q^a dq^i) +\Omega.
\label{sim}
\end{equation}
 To achieve the
notational simplifications, we introduce $\eta^I=(p_i, q^j)$ and
$x^M=(\xi^A,\bar\xi^B,\eta^I)$. $\xi$'s  and $\bar\xi$'s are
the internal coordinates. Then we can write $\Omega_T=
\frac{1}{2}\Omega_{MN}dx^M\wedge dx^N$.
Using Eq. (\ref{sim}), one finds the following inverse
matrix $\Omega^{MN}$:
\begin{equation}
\Omega^{MN}=\left( \begin{array}{cc}
    \Omega^{AB}  & -F^{KJ}\Omega^{AC}A^a_K(\partial Q^a/\partial \xi^C) \\
F^{KI}\Omega^{BD}A^a_K(\partial Q^a/\partial\xi^D) & F^{IJ}
\end{array} \right)
\label{mini1}
\end{equation}
where $F^{IJ}$ is the inverse matrix of
 $F_{IJ}\equiv \omega_{IJ}-f_{abc}A^a_IA^b_JQ^c$
and is given by
\begin{equation}
 F^{IJ}=\left(\begin{array}{cc}
   F^a_{ij}Q^a & -I  \\
             I & 0 \end{array} \right).
\label{def4}
\end{equation}
Here $F^a_{ij}\equiv \partial_jA^a_i-\partial_iA^a_j-f^a_{\ bc}A^b_iA^c_j$
is the Yang-Mills field strength.

The Poisson bracket on ${\cal P}$  is defined by the use of
inverse matrix $\Omega^{MN}$ as before
\begin{equation}
\{F,H\}=\Omega^{MN}\frac{\partial F}{\partial x^M}
\frac{\partial H}{\partial x^N}.
\end{equation}
And we find the following Poisson brackets along with
$\{Q^a, Q^b\}=-f^{ab}_{\ \ c}Q^c$,
\begin{equation}
\{Q^a,p_i\}=-f^a_{\ bc}A^b_iQ^c, \ \ \ \ \ \ \ \ \{Q^a,q^i\}=0
\end{equation}
\[ \{p_i, p_j\}=F^a_{ij}Q^a, \ \ \ \{p_i,q^j\}=-\delta_i^j,
 \ \ \ \{q^i,q^j\}=0.\]
The above relations are in accordance with the minimal substitution
\begin{equation}
p_i\rightarrow P_i=p_i - A_i^a Q^a.
\end{equation}
In terms of canonical momentum $P_i$, we have, among others,
\begin{equation}
\{Q^a,P_i\}=0 \ \ \ \ \{P_i, P_j\}=0. \ \ \ \ \{P_i,q^j\}=-\delta_i^j.
\label{cano}
\end{equation}
Thus, one can work in $(p_i,q^i,Q^a)$
coordinates using the symplectic
structure given by Eqs. (\ref{mini1}) and (\ref{def4}) or with
 $(P_i,q^i,Q^a)$ using the
canonical symplectic structure without mixing between
 $ P_i$ and $Q^a$.
The two procedures are equivalent\cite{ster}.

Consider, for example, the free Hamiltonian $H=(1/2m) p^2$
 with the symplectic
structure given by Eqs. (\ref{mini1}) and (\ref{def4}).
 The Hamiltonian equations of motion
\begin{equation}
\dot x^M=\Omega^{MN}\frac{\partial H}{\partial x^N}
\end{equation}
reproduce the well known Wong's equations \cite{wong}
\begin{equation}
m\ddot{q}_i =  F^a_{ij}Q^a\dot{q}^j \quad
\dot{Q}^a = -f^a_{\ bc} A^b_i \dot{q}^i Q^c \label{eul2}
\end{equation}
which describe the dynamics of an isospin particle in an external
gauge fields $A_i^a$.
Minimal substitution implies that alternatively, we can work with
\begin{equation}
 H =  {1\over 2 m}\left(P_i-A^a_i Q^a\right)^2\label{hamil}
\end{equation}
with canonical symplectic structure Eq. (\ref{cano}).
Obviously, we get the same equations of motions.
The above procedures can be generalized to a system of many particles
in an obvious manner and
can be applied to a system of NACS particles.
 We end up with the Hamiltonian
Eq. (\ref{mhami}) where the $Q^a$'s are now given by $SU(N)$ isospin
functions (\ref{isof}) on each  ${\cal O}_{\{n_1,n_2,\cdots,n_l\}}$.
Also, $SU(N)$ Gauss law constraint and its solution in complex
spatial coordinates
 are in the same form as
the equations (\ref{gauss}) and (\ref{sol}) with
$SU(N) \  Q^a$'s.
Then, the quantum mechanical Hamiltonian of
a system of NACS particles is obtained in the same expression as the
equation (\ref{ham}) with the isospin operators
satisfying $SU(N)$ algebra;
$[\hat Q^a_\alpha,\hat Q^b_\alpha]=
if^{ab}_{\ \ c}\hat Q^c_\alpha\delta_{\alpha\beta}$.
We can infer that most of the quantum mechanical properties of
 $SU(2)$ NACS particles carry qualitatively over to
$SU(N)$ case and in particular,
  a system of $SU(N)$ NACS particles also exhibit
$SU(N)$ braid statistics described by $SU(N)$ KZ equation.

It is worth mentioning the origin of the Gauss
law constraint (\ref{gauss}) in our Hamiltonian formulation at this point.
It can be shown \cite{oh3} that it is the condition
of the vanishing momentum map of  gauge transformations
in the total phase space in which the phase space of gauge connection
is also included. In our Hamiltonian approach, we neglected the
space of gauge connection for simplicity and reduced phase space
of a system of $N_p$ NACS particles is given by
an associated bundle
$\prod_\alpha {\cal P}_\alpha\equiv \prod_\alpha T^*M_\alpha\times
{\cal O}^\alpha_{\{n_1,n_2,\cdots,n_l\}}$
($\alpha=1,\cdots,N_p$) with the gauge connection given by
the KZ connection, Eq. (\ref{sol}) with $P(z)=0.$
 When the NACS particles are indistinguishable,
configurations that differ by the interchange of two particles
must be identified and the phase space is given by \cite{wuu}
\begin{equation}
T^*\left[\frac{\prod_\alpha M_\alpha-{\cal D}}{S_{N_p}}\right]
\times {\cal O}^\alpha_{\{n_1,n_2,\cdots,n_l\}}.
\label{identical}
\end{equation}
In the above equation,
${\cal D}$ is the set of points where ${\bf q}_\alpha=
{\bf q}_\beta$ for some $\alpha,\beta$ and
$S_{N_p}$ is the permutation group of $N_p$ objects.

\def\theequation{\arabic{section}.\arabic{equation}}
\section{Coherent State Quantization}
\setcounter{equation}{0}
In this section, we quantize the Hamiltonian (\ref{mhami})
with $Q^a$'s given by the isospin functions (\ref{isof})
in the coherent state quantization method \cite{klau}.
This method is used only for the internal degrees of freedom
for convenience.
The external gauge field $A^a_i$ will be arbitrary
for the time being. Later, when we consider the NACS particles,
it will be substituted by the KZ connection (\ref{sol}).
Let us consider the propagator
\begin{equation}
K_{FI}=<q_F,\bar \xi_F\vert e^{-i\hat H(t_F-t_I)}\vert q_I,\xi_I>
\label{propagator}
\end{equation}
where $\vert q_I,\xi_I>=\prod_{\alpha=1}^N\vert {\bf q}_{\alpha I}>
\vert \xi^i_{\alpha I}>$ and similarly for $<q_F,\bar \xi_F\vert$.
$\xi^i_\alpha$ is the internal complex coordinate on the coadjoint
orbit of $\alpha$-th particle ${\cal O}^\alpha_{\{n_1,n_2,\cdots,n_l\}}
=SU(N)/SU(n_1)\times\cdots\times SU(n_l)\times U(1)^{l-1}$.
$\hat H$ is the quantum mechanical operator of the
Hamiltonian (\ref{mhami}).
Finally, $\vert \xi>$ is the generalized coherent
state \cite{klau} defined by
\begin{equation}
\vert \xi>=\exp (\xi\cdot E)\vert\Lambda>.
\label{definition}
\end{equation}
$\Lambda$ is the highest weight which can be expressed as
\begin{equation}
\Lambda=\sum_s \mu_sf_s.
\end{equation}
Here, $\mu_s$ is a non-negative integer and $f_s$ is the highest
weight of the fundamental representation. The summation over $s$
is  done in such a way that that the maximum stability group
of $\vert\Lambda>$ is $SU(n_1)\times\cdots\times SU(n_l)\times U(1)^{l-1}$
and the corresponding geometry of coherent state is
$SU(N)/SU(n_1)\times\cdots\times SU(n_l)\times U(1)^{l-1}$.
The existence of such a weight is guaranteed by the Borel-Weil-Bott
theorem \cite{helg}.
The normalization for the coherent state is chosen
for convenience in the following manner
\begin{equation}
<\bar \xi^\prime\vert\xi>=e^{W(\bar\xi^\prime,\xi)}
\label{normalization}
\end{equation}
where $W$ is the K\"ahler potential on the orbit
${\cal O}_{\{n_1,n_2,\cdots,n_l\}}.$ Note that with this
normalization in Eq. (\ref{definition}), $\vert\xi>$ is
a holomorphic function of $\xi$ and $<\bar\xi\vert=(\vert\xi>)^\dagger$
an antiholomorphic state. Then, $W(\bar\xi^\prime,\xi)$ is
holomorphic in $\xi$ and antiholomorphic in $\xi^\prime$.
The resolution of unity
for the coherent state is expressed as
\begin{equation}
I=\int d\mu(\bar\xi,\xi)\vert\xi><\bar\xi\vert e^{-W(\bar\xi,\xi)}.
\label{identity}
\end{equation}
Also we have $I_q=\int dq \vert q><q\vert, I_p=\int dp\vert p><p\vert$.
We will not sometimes write  bold faces for $p,q$
and particle indices $\alpha$ and $\beta$ unless confusion arises.

We first perform the lattice evaluation of the propagator. Divide the time
$T\equiv t_F-t_I$ into $\bar N+1$  steps of equal length
$\epsilon$ so that $(\bar N+1)\epsilon=T$,
$t_1=t_I$ and $t_{\bar N+1}=t_F$. The boundary value is given by
$\bar\xi(t_{\bar N+1})=\bar\xi_F$ and $\xi(t_1)=\xi_I$. Inserting the
resolution of unity $I\times I_q$ and writing $\xi(t_n)\equiv \xi(n)$,
we have
\begin{eqnarray}
K_{FI}=\int\cdots\int\prod_{n=1}^{\bar N}&&d\mu(\bar\xi(n),\xi(n))dq(n)
e^{-W(\bar\xi(n),\xi(n))}\prod_{n=1}^{\bar N+1}<\bar\xi(n)\vert\xi(n-1)>
\times\nonumber\\
&&\left[<q(n)\vert q(n-1)>
-i\epsilon \frac{<q(n)\bar\xi(n)\vert \hat H\vert q(n-1)\xi(n-1)>}
{<\bar\xi(n)\vert\xi(n-1)>}\right].
\end{eqnarray}
Using the kernel $<\bar\xi(n)\vert\xi(n-1)>=\exp(W(\bar\xi(n),\xi(n-1)))$,
$\bar\xi(n)=\bar\xi(n-1)+d\xi(n-1)$ and
treating the space part in the standard manner by inserting $I_p$ repeatedly,
we have in the continuum limit $(\epsilon\rightarrow 0)$
\begin{equation}
K_{FI}=C\int d\mu(\bar\xi,\xi)dpdq e^{-\log W(\bar\xi_F,\xi_F)}
e^{i\int_{t_I}^{t_F}Ldt}
\label{desire}
\end{equation}
where the Lagrangian is given by
\begin{equation}
L=\sum_\alpha ({\bf p}_\alpha
\cdot\dot{\bf q}_\alpha -i\frac{\partial W(\bar\xi_\alpha,\xi_\alpha)}
{\partial \bar\xi_\alpha}\cdot\dot{\bar\xi}_\alpha
)-H.
\label{pathintegral}
\end{equation}
The Hamiltonian is given by
\begin{equation}
H=\sum_\alpha\frac{<q_\alpha,\bar\xi_\alpha\vert\hat H
\vert q_\alpha,\xi_\alpha>}
{<\bar\xi_\alpha\vert\xi_\alpha>}
\end{equation}
with $\hat H$ given by the operator form of Eq. (\ref{mhami})
\begin{equation}
\hat H = \sum_\alpha {1\over 2 m_\alpha}\left(\hat p^i_\alpha-A^{ai}
(\hat{\bf q}_\alpha)
\hat Q^a_\alpha\right)^2.\label{qmhami}
\end{equation}
It is to be noticed that in the above $\hat H$ the operator
$\hat Q^a_\alpha$ is the coherent state
representation expressed in terms of
the internal coordinates $\xi_\alpha$'s of the coadjoint orbit
${\cal O}^\alpha_{\{n_1,n_2,\cdots,n_l\}}$. We are interested in
the  differential operator representation satisfying
$[\hat Q^a_\alpha,\hat Q^b_\alpha]=
if^{ab}_{\ \ c}\hat Q^c_\alpha\delta_{\alpha\beta}$ and assume
such a representation is possible.

Using the complex coordinates  for spatial part and KZ
connection given in Eq. (\ref{sol}) with $P(z)=0$, we again
recover the quantum mechanical model given by Eq. (\ref{ham})
where  the isospin operators $\hat Q^a_\alpha$'s satisfy the $SU(N)$
algebra: $[\hat Q^a_\alpha,\hat Q^b_\alpha]=
if^{ab}_{\ \ c}\hat Q^c_\alpha\delta_{\alpha\beta}$.
But as mentioned just before, $\hat Q^a_\alpha$'s are now a differential
operator and the wave function is now function of both
spatial coordinates and internal coordinates: $\Psi\equiv
\Psi(z_1,\cdots,z_{N_p}, \xi^i_1,\cdots,\xi^i_{N_p},
\bar\xi^i_1,\cdots,\bar\xi^i_{N_p})$.
The $\bar z$ dependence is dropped for convenience. It is
a single component wave function.
It is easy to show that the KZ equation can  be
written as follows:
\begin{equation}
\frac{\partial U^{-1}}{\partial z_\alpha}+
\frac{1}{2\pi\kappa}\sum_{\beta\neq \alpha}
\frac{\hat Q^a(\xi_\alpha,\bar\xi_\alpha)
\hat Q^a(\xi_\beta,\bar\xi_\beta)}{z_\alpha-z_\beta}
U^{-1}=0
\label{kzeq}
\end{equation}
 where $\hat Q^a(\xi_\alpha,\bar\xi_\alpha)$ is
the differential operator of $\hat Q^a_\alpha$ in the
coherent state representation. The holomorphicity of the state
$<\xi\vert=<\Lambda\vert \exp (\xi\cdot E^\dagger)$
enables one to choose the holomorphic
polarization of $U^{-1}(z,\xi)\equiv <\xi\vert U^{-1}(z)>
=\int d\mu(\bar\xi^\prime,\xi^\prime)
e^{W(\xi,\bar\xi^\prime)-W(\bar\xi^\prime,\xi^\prime)}
<\xi^\prime\vert U^{-1}(z)>$ which is obviously holomorphic
in $\xi$. Note that this choice is possible, because we have chosen
the normalization given by the Eq. (\ref{normalization}).
{}From now on, we will be working on the holomorphic
polarization:  $U^{-1}\equiv
U^{-1}(z_1,\cdots,z_{N_p}, \xi^i_1,\cdots,\xi^i_{N_p})$.
Also, the wave function is a holomorphic function:
$\Psi\equiv
\Psi(z_1,\cdots,z_{N_p}, \xi^i_1,\cdots,\xi^i_{N_p})$.
The inner product Eq. (\ref{inner}) is modified into
\begin{equation}
<\Psi_1 |\Psi_2> = \int \prod_\alpha
dz_\alpha d\bar z_\alpha  d\mu(\bar\xi_\alpha,\xi_\alpha)
 \bar\Psi_1(z,\bar\xi) U^\dagger
(z,\bar\xi) U(z,\xi) \Psi_2 (z,\xi)e^{-W(\bar\xi,\xi)}.
\end{equation}
 The Hamiltonian  Eq. (\ref{ham}) with $SU(N)\ \ \hat Q^a_\alpha$'s
 is hermitian with respect to
this inner product
assuming the hermiticity of $\hat Q^a(\xi_\alpha,\bar\xi_\alpha)$.

We discuss about $CP(N-1)$ case in details. The K\"ahler
potential $W$ is given by
\begin{equation}
W(\bar\xi, \xi)=\sum_\alpha J_\alpha
\log(1+\bar\xi_\alpha\xi_\alpha)
 \end{equation}
for some integer $J_\alpha$. The Lagrangian is expressed as
\begin{equation}
L=\sum_\alpha\left(p^{\bar
z}_\alpha \dot z_{\alpha}+ p^{z}_\alpha \dot{\bar z_\alpha}-
iJ_\alpha\frac{\xi_\alpha\cdot\dot{\bar\xi_\alpha}}
{1+\vert\xi^\alpha\vert^2}\right)-H\label{path}
\end{equation}
with the Hamiltonian $H$ given by Eq. (\ref{prop}) with the isospin
functions being given by the Eq. (\ref{deff2})
\begin{equation}
Q^a(\xi_\alpha,\bar\xi_\alpha)=i\sum_\alpha\sum_{I,K=0}^{N-1}
J_\alpha {\bar \Xi}_{\alpha I}T^a_{IK}\Xi_{\alpha K}\label{def2}
\end{equation}
where $\Xi_{\alpha 0}=\frac{1}{\sqrt{1+\mid\xi_\alpha\mid^2}}$ and
$\Xi_{\alpha i}=\frac{\xi_\alpha^i}{\sqrt{1+\vert\xi_\alpha\vert^2}}$ are
substituted into the final expression. The quantum mechanical
Hamiltonian is again given by the Eq. (\ref{ham}) but with
 the quantum mechanical isospin operator being expressed by
\begin{equation}
\hat Q^a(\xi_\alpha)=i[T^a_{i0}+T^a_{ij}\xi^j_\alpha-
T^a_{00}\xi^i_\alpha-T^a_{0j}\xi^j_\alpha\xi^i_\alpha]
\frac{\partial}{\partial\xi^i_\alpha}+
iJ_\alpha T^a_{00}+iJ_\alpha T^a_{0i}\xi^i_\alpha.
\label{cpop}
\end{equation}
The above differential operator satisfy $[\hat Q^a(\xi_\alpha),
\hat Q^b(\xi_\beta)]
=if^{ab}_{\ \ c}\hat Q^c(\xi_\alpha)\delta_{\alpha\beta}$.

The explicit form of the differential KZ equation (\ref{kzeq})
in this representation can be given: for example, in $SU(2)$ case,
\begin{equation}
\frac{\partial U^{-1}}{\partial z_\alpha}+
 {1\over 2\pi\kappa}\sum_{\beta\neq \alpha}
\frac{\frac{1}{2}(\hat Q^+(\xi_\alpha) \hat Q^-(\xi_\beta)+
\hat Q^-(\xi_\alpha)\hat Q^+(\xi_\beta))+\hat Q^3
(\xi_\alpha)\hat Q^3(\xi_\beta)}{z_\alpha-z_\beta}U^{-1}=0\label{dkzeq}
\end{equation}
where we defined
$\hat Q^\pm(\xi_\alpha)=\hat Q^1(\xi_\alpha)\pm i\hat Q^2(\xi_\alpha)$
and they are given by
\begin{equation}
\hat Q^+(\xi_\alpha)=\xi^2_\alpha\frac{\partial}{\partial\xi_\alpha}-
J_\alpha\xi_\alpha,\ \ \hat Q^-(\xi_\alpha)=-\frac{\partial}
{\partial\xi_\alpha},\ \ \hat Q^3(\xi_\alpha)=\xi_\alpha\frac{\partial}
{\partial\xi_\alpha}-\frac{J_\alpha}{2}.
\end{equation}
We see that the above is the $SU(2)$ generalization of the Bargmann
representation \cite{klau}. For example,
the $(J+1)$-dimensional  irreducible representation of the operator
is given by the holomorphic polynomial of order $J$:
$\psi_J(\xi)=\sum_{n=0}^Ja_n\xi^n$. The highest weight state is given by
$\psi_{JJ}(\xi)=a_J\xi^J$ and they satisfy $\hat Q^+(\xi)\psi_{JJ}(\xi)=0,\
\hat Q^-(\xi)\psi_{JJ}(\xi)=-J\psi_{JJ-1}(\xi),\
 \hat Q^3(\xi)=(J/2)\psi_{JJ}(\xi).$ We also have
$\hat Q^a(\xi)\hat Q^a(\xi)=(J/2)(J/2 +1)$. Note the correspondence
with the matrix representation: the usual angular momentum $j=J/2,\
m=M-(J/2),\ \mbox{and} \ \ \psi_{JM}\rightarrow\vert j,m>$.

It is to be noted that antiholomorphic representation is given by
the complex conjugation of the above equation,
$\hat Q^{a*}(\xi_\alpha)$'s, and they satisfy
$[\hat Q^{a*}(\xi_\alpha),
\hat Q^{b*}(\xi_\beta)]
=-if^{ab}_{\ \ c}\hat Q^{c*}(\xi_\alpha)\delta_{\alpha\beta}$.
Also, the operator (\ref{cpop}) is Hermitian with respect to the
following inner product:
\begin{equation}
<\psi_1\vert\psi_2>=\int  d\mu(\bar\xi,\xi)
\bar\psi_1(\bar\xi)\psi_2(\xi)e^{-W(\bar\xi,\xi)}.
\end{equation}
Finally, it can be easily checked that
\begin{equation}
<\bar\xi_\alpha\vert \hat Q^a_\alpha\vert\xi_\alpha>
\equiv <\bar\xi_\alpha\vert \hat Q^a_\alpha\vert
\xi_\alpha^\prime>\Big\vert_{\xi_\alpha^\prime\rightarrow\xi_\alpha}=
(\hat Q^a(\xi^\prime_\alpha)<\xi^\prime_\alpha\vert\bar\xi_\alpha>)^*
\Big\vert_{\xi_\alpha^\prime=\xi_\alpha}
=Q^a(\xi_\alpha,\bar\xi_\alpha)<\bar\xi_\alpha\vert\xi_\alpha>
\end{equation}
using the reproducing kernel of $CP(N-1)$
\begin{equation}
<\bar\xi_\alpha^\prime\vert\xi_\alpha>
=(1+\bar\xi_\alpha^\prime\cdot\xi_\alpha)^{J_\alpha}.
\end{equation}

Now let us discuss the possible solution
of the KZ equation in the $CP(N-1)$ case
with the differential operator given by (\ref{cpop}).
For our purpose, we rewrite the KZ equation (\ref{kzeq})
in the following integral differential equation
\begin{eqnarray}
U^{-1}(z_1,\cdots,z_{N_p}; \xi^i_1,\cdots,\xi^i_{N_p})&=&
U^{-1}_0(\xi^i_1,\cdots,\xi^i_{N_p})-\frac{1}{2\pi\kappa}
\int_{\Gamma}\sum_\alpha
dz^\prime_\alpha\sum_{\beta\neq\alpha}
\frac{1}{z^\prime_\alpha-z^\prime_\beta}\times
\nonumber\\
&\times&\hat Q^a(\xi_\alpha)\hat Q^a(\xi_\beta)
U^{-1}(z^\prime_1,\cdots,z^\prime_{N_p};\xi^i_1,\cdots,\xi^i_{N_p})
\label{kzeqint}
\end{eqnarray}
where $\Gamma$ is a path in the $N_p$-dimensional
complex space with one end point fixed and the other
being $z_f=(z_1,\cdots,z_{N_p})$.
$U^{-1}_0(\xi^i_1,\cdots,\xi^i_{N_p})$ is independent of
spatial coordinates and necessary to take care of the
boundary condition.   It is well known that
the exact solution can be achieved in the two body case.
It can be easily inferred that the solution is given by
\begin{equation}
U^{-1}(z_1,z_2; \xi^i_1,\xi^i_2)
=\exp\left(-\frac{1}{2\pi\kappa}(\log(z_1-z_2)+c_{12})
\hat Q^a(\xi_1)\hat Q^a(\xi_2)\right)
U^{-1}_0(\xi^i_1,\xi^i_2)\label{exact}
\end{equation}
where $c_{12}$ is a constant which is inserted to adjust the boundary
condition. An exact expression can be obtained for a couple
of cases. Following the discussion of $SU(2)$ Bargmann representation,
let us consider the case in which we have the Clebsch-Gordan series
\begin{equation}
U^{-1}_0(\xi^i_1,\xi^i_2)=\psi_{JM}=\sum_{M_1,M_2}C(\frac{J_1}{2}
M_1-\frac{J_1}{2},
\frac{J_2}{2}M_2-\frac{J_2}{2};\frac{J}{2}M-\frac{J}{2})
\psi_{J_1M_1}\psi_{J_2M_2}
\end{equation}
where $C(\frac{J_1}{2}m_1,\frac{J_2}{2}m_2;\frac{J}{2}m)$
is the Clebsch-Gordan coefficients. Then it can be easily shown that
$\hat Q^a(\xi_1)\hat Q^a(\xi_2)U^{-1}_0(\xi^i_1,\xi^i_2)=(1/2)
[\frac{J}{2}(\frac{J}{2}+1)-\frac{J_1}{2}(\frac{J_1}{2}+1)-
\frac{J_2}{2}(\frac{J_2}{2}+1)]U^{-1}_0(\xi^i_1,\xi^i_2)$ and the solution
is given by the Eq. (\ref{exact}) with $\hat Q^a(\xi_1)\hat Q^a(\xi_2)$
replaced by $(1/2)[\frac{J}{2}(\frac{J}{2}+1)-\frac{J_1}{2}(\frac{J_1}{2}+1)-
\frac{J_2}{2}(\frac{J_2}{2}+1)]$.
The extension to general $CP(N-1)$ is immediate and will be reported elsewhere.
Another case of interest would be
the one in which $U^{-1}_0(\xi^i_1,\xi^i_2)$ is given by the coherent state
\begin{equation}
U^{-1}_0(\xi^i_1,\xi^i_2)=<\xi_1\vert\bar\zeta_1><\xi_2\vert\bar\zeta_2>
\end{equation}
for some $\vert\bar\zeta_1>$ and $\vert\bar\zeta_2>$.
The solution can be written as (with $c_{12}=0$
for convenience)
\begin{equation}
U^{-1}(z_1,z_2; \xi^i_1,\xi^i_2)
=\sum_{n=0}^\infty(\frac{-\log(z_1-z_2)}{2\pi\kappa})^n
(\hat Q^a(\xi_1)\hat Q^a(\xi_2))^n
(1+\xi_1\cdot\bar\zeta_1)^{J_1}
(1+\xi_2\cdot\bar\zeta_2)^{J_2}.
\end{equation}
 We see that in the $SU(2)$ case, for example,
in the large $\kappa$ limit neglecting terms of order
$(1/\kappa)^2$, $U^{-1}(z_1,z_2; \xi^i_1,\xi^i_2)$
is a polynomial of order $J_1$ and $J_2$
in $\xi_1$ and $\xi_2$ respectively. It would be interesting if
one could find general solutions of Eq. (\ref{kzeqint}).

\section{Conclusion}
\setcounter{equation}{0}
In this paper, we investigated in detail the classical and quantum aspect
of a system of $SU(N)$ NACS particles.
We discussed about the most general
phase space of $SU(N)$ internal degrees of
freedom which can be identified as one of the coadjoint orbits
of $SU(N)$ group by the method of symplectic reduction.
A detailed constraint analysis on each orbit by Dirac method was given.
The quantum aspect of the theory
was explored by using the method of the coherent state for the
internal degrees of freedom. Coherent state corresponding
to the geometry of each coadjoint orbit was introduced and
an explicit path integral representation was derived.
 Especially, a coherent state representation
of the KZ equation was given and
possible solutions in this representation were discussed.

There remain several topics to be discussed further in this approach.
First of all, it would be interesting if the constraint analysis
and coherent state quantization approach of this paper could be
generalized to  arbitrary groups including the non-compact ones
and applied to give an explicit construction
of the corresponding Darboux variables on each coadjoint orbit
 as was done on the maximal orbits of unitary and orthogonal
group \cite{alek2}. The results could give the functional integral
quantization of spin \cite{niel} on the most general coadjoint orbits.

We performed coherent state quantization of NACS particles and obtained
a path integral representation of NACS particles in Section 4. For example,
by using  the  complex spatial
coordinate and substituting the KZ connection (\ref{sol})
with $P(z)=0$ in the holomorphic gauge into Eq. (\ref{pathintegral}),
we obtain the desired propagator (\ref{desire}).
Since the NACS particles are the non-Abelian generalizations
of anyons and the propagator of a system of indistinguishable anyons
 is a representation of the braid
group \cite{wu}, our propagator should also provide
a non-Abelian generalization of
path integral representation of the braid group
on the phase space Eq. (\ref{identical}). It could be called the
coherent state representation of the braid group. The detailed
analysis of a system of indistinguishable NACS particles will be
reported elsewhere \cite{philoh}.

Note that in the usual expression of KZ equation (\ref{mkzeq}),
the braid operator or monodromy
$\exp(i\hat Q^a_\alpha \hat Q^a_\beta/\kappa)$ is given
as a matrix representation whereas in the coherent state approach
this is an holomorphic differential operator.
 The relation between the two approach
is connected by the simple exchange of the usual angular
momentum basis and coherent state.
This could have two applications.
First, we recall that the difficulty in finding the solution
of the KZ equation in matrix approach
lies in the non-existence of the common eigenvectors of
the braid operators in general \cite{kapu}.  This difficulty
may be cured in our approach because it is replaced with finding
the possible solutions of the KZ differential equation.
Second, since the braid operator satisfies the Yang-Baxter equation
and exhibits the non-Abelian braid statistics \cite{kohn}, our approach
could give a new interpretation of the Yang-Baxter equation.
The possibility of exact solutions and
detailed property of non-Abelian braid statistics in this approach will
be reported elsewhere.

Finally, it may be interesting if the generalized Bargmann representation
can be explicitly obtained for other coadjoint orbit
 as well and all the holomorphic irreducible representations are
 explicitly calculated. For example, for the maximal orbit,
the Bruhat coordinatization \cite{pick}
can be used in the construction of the coherent state
and an explicit holomorphic representation of
$\hat Q^a(\xi)$ can be obtained by using the method of
geometric quantization \cite{wood}. The
details will be reported elsewhere \cite{philoh}.

\acknowledgments
I would like to thank Professors M.-H. Kim, J. Klauder, P. L. Robinson
for useful discussions
and Dr. N. P. Landsman for showing interest in this work.
This work was supported by  the KOSEF
through the project number 95-0702-16-3 and  C.T.P. at S.N.U.
and by the  Ministry of Education through the Research Institute of
Basic Science.
It was also supported by Seok-Cheon research grant from Sung Kyun Kwan
university.

\end{document}